\documentclass[aps, prl, twocolumn, showpacs, superscriptaddress]{revtex4}

\usepackage{graphicx}
\usepackage{dcolumn}
\usepackage{bm}

\begin{document}

\title{Absence of static magnetic order in lightly-doped Ti$_{1-x}$Sc$_x$OCl down to 1.7~K}

\author{A.A. Aczel}
\altaffiliation{author to whom correspondences should be addressed: E-mail:[aczelaa@ornl.gov]}
\affiliation{Department of Physics and Astronomy, McMaster University, Hamilton, ON, Canada, L8S 4M1}
\affiliation{Neutron Scattering Science Division, Oak Ridge National Laboratory, Oak Ridge, TN 37831, USA}
\author{G.J. MacDougall}
\affiliation{Department of Physics and Astronomy, McMaster University, Hamilton, ON, Canada, L8S 4M1}
\affiliation{Neutron Scattering Science Division, Oak Ridge National Laboratory, Oak Ridge, TN 37831, USA}
\author{F.L. Ning}
\affiliation{Department of Physics and Astronomy, McMaster University, Hamilton, ON, Canada, L8S 4M1}
\affiliation{Department of Physics, Columbia University, New York, NY 10027, USA}
\author{J.A. Rodriguez}
\affiliation{Department of Physics and Astronomy, McMaster University, Hamilton, ON, Canada, L8S 4M1}
\affiliation{Laboratory for Muon Spin Spectroscopy, Paul-Scherrer Institute, CH-5232 Villigen PSI, Switzerland}
\author{S.R. Saha}
\affiliation{Department of Physics and Astronomy, McMaster University, Hamilton, ON, Canada, L8S 4M1}
\affiliation{Center for Nanophysics and Advanced Materials, Department of Physics, University of Maryland, College Park, MD 20742, USA}
\author{F.C. Chou}
\affiliation{Center for Condensed Matter Sciences, National Taiwan University, Taipei 106, Taiwan}
\author{T. Imai}
\affiliation{Department of Physics and Astronomy, McMaster University, Hamilton, ON, Canada, L8S 4M1}
\affiliation{Canadian Institute of Advanced Research, Toronto, Ontario, Canada, M5G 1Z8}
\author{G.M. Luke}
\affiliation{Department of Physics and Astronomy, McMaster University, Hamilton, ON, Canada, L8S 4M1}
\affiliation{Canadian Institute of Advanced Research, Toronto, Ontario, Canada, M5G 1Z8}

\date{\today}

\begin{abstract}
Impurity-induced magnetic order has been observed in many quasi-1D systems including doped variants of the spin-Peierls system CuGeO$_3$. TiOCl is another quasi-1D quantum magnet with a spin-Peierls ground state, and the magnetic Ti sites of this system can be doped with non-magnetic Sc. To investigate the role of non-magnetic impurities in this system, we have performed both zero field and longitudinal field $\mu$SR experiments on polycrystalline Ti$_{1-x}$Sc$_x$OCl samples with x~$=$~0, 0.01, and 0.03. We verified that TiOCl has a non-magnetic ground state, and we found no evidence for spin freezing or magnetic ordering in the lightly-doped Sc samples down to 1.7 K. Our results instead suggest that these systems remain non-magnetic up to the x~$=$~0.03 Sc doping level.
\end{abstract}


\maketitle

\section{\label{sec:level1}Introduction}
Low dimensional magnets are of current interest, due to their possible relevance to high temperature superconductivity and their penchant for possessing exotic ground states\cite{05_lake, 99_kageyama, 96_dagotto, 10_hsu}. One subgroup of these materials is the spin-Peierls (SP) systems, which possess non-magnetic spin-singlet ground states. These materials consist of quasi-1D Heisenberg antiferromagnetic chains that dimerize at low temperatures due to strong magnetoelastic coupling. There are currently three known inorganic SP systems: CuGeO$_3$\cite{93_hase}, TiOCl\cite{03_seidel}, and TiOBr\cite{05_sasaki}. Organic SP systems such as TTF-CuS$_4$C$_4$(CF$_3$)$_4$\cite{75_bray} and MEM-(TCNQ)$_2$\cite{79_huizinga} were actually discovered first, but these are much more difficult to study due to very low magnetic moment densities.

The quasi-1D magnet TiOCl consists of magnetic Ti$^{3+}$ chains (spin S $= \frac{1}{2}$ in 3d$^1$ state), and undergoes successive phase transitions at T$_{c2}$ = 94 K to an incommensurate SP phase and T$_{c1}$ = 66 K to the commensurate SP ground state\cite{03_imai}. The presence of two phase transitions is in contrast to conventional SP systems, where only one transition is observed. NMR\cite{03_imai} and x-ray scattering measurements\cite{05_shaz, 07_clancy, 07_abel} find a uniform dimerization below T$_{c1}$ along the b-axis, providing evidence for commensurate SP behaviour. 

The nature of the incommensurate SP phase between T$_{c1}$ and T$_{c2}$ is currently not well understood. Magnetic susceptibility and NMR measurements provide evidence that the upper transition is associated with the onset of dimerization and the opening of a spin gap below T$_{c2}$\cite{03_seidel, 03_imai}. Recent x-ray measurements have also found an incommensurate lattice distortion along the a and b-axes which is quite long-ranged ($>$~2000~\AA~in each case)\cite{07_abel}. Finally, $^{35}$Cl NMR measurements detect two peaks in the frequency spectrum corresponding to the I$_z = -\frac{1}{2}$ to $\frac{1}{2}$ central transition. One of these peaks is well-defined but the other is much broader\cite{07_saha}. These experiments suggest that while there is dimerization in this incommensurate phase, the Ti-Ti intradimer distance is not constant and adjacent Ti chains have a small relative shift to one another along the chain direction. This small relative shift vanishes below T$_{c1}$, where the ``lock-in" transition occurs. At this point, all Ti chains are aligned with one another and the Ti-Ti intradimer distance in a given chain is constant. Complicating the picture further in the incommensurate phase regime are recent x-ray measurements that have detected commensurate fluctuations coexisting with incommensurate Bragg peaks\cite{07_clancy}.

It has long been appreciated that substituting a small amount of non-magnetic impurities for the magnetic sites can lead to long-range magnetic order in conventional SP systems\cite{95_hase}, and so doping studies of these materials have been of great interest. In particular, it is possible to dope the magnetic Cu sites in CuGeO$_3$ with non-magnetic Zn$^{2+}$\cite{93_hase, 95_oseroff, 97_kojima, 97_martin, 98_grenier, 98_manabe}, Mg$^{2+}$\cite{98_grenier}, or Cd$^{2+}$\cite{07_haravifard, 98_lumsden}, as well as magnetic Ni$^{2+}$\cite{95_oseroff, 98_grenier}, Co$^{2+}$\cite{97_anderson}, or Mn$^{2+}$\cite{95_oseroff}. Systems with Si$^{4+}$\cite{95_oseroff, 97_kojima} doped in for Ge$^{4+}$ have also been created. These materials have generally revealed phase diagrams with some common features, including the loss of SP order at a critical doping concentration x$_c$, a ``dimerized antiferromagnetic ground state" for the lightly-doped compounds with x~$<$~x$_c$, and a uniform antiferromagnetic ground state for systems with x~$>$~x$_c$\cite{02_uchinokura}. In addition, a detailed study on the Cu$_{1-x}$Zn$_x$GeO$_3$ systems indicated the presence of impurity-induced long-range magnetic order down to the lowest doping concentration studied (x~$=$~0.001)\cite{98_manabe}. This suggests the absence of a required critical doping concentration to achieve magnetic order and is in agreement with theoretical work\cite{96_fukuyama}. 

One exception to these properties was found in the case of Cu$_{1-x}$Cd$_x$GeO$_3$, where long-range magnetic order was not observed for x~$\le$~0.002\cite{07_haravifard} and the universality class was found to change upon doping from three-dimensional XY to mean-field\cite{98_lumsden}. These features were suggested to be consequences of local strain fields induced by the presence of larger dopant ions, as the ionic radius of Cd$^{2+}$ (0.97~\AA) is much larger than that of Cu$^{2+}$ (0.72~\AA). For all other doped systems investigated, the ionic radius of the dopant ion is either smaller or comparable to that of the ion being replaced. 

Although many detailed studies on doped CuGeO$_3$ have been performed, very little is currently known regarding how dopants affect the unconventional SP ground state of TiOCl. An early report on this topic discussed susceptibility results of Ti$_{1-x}$Sc$_{x}$OCl\cite{03_seidel}. Note that substituting non-magnetic Sc$^{3+}$ for magnetic Ti$^{3+}$ is essentially analogous to substituting a non-magnetic ion for Cu$^{2+}$ in CuGeO$_3$. In both cases, the non-magnetic ions should lead to the destruction of some dimers and create quasi-free spins. Accordingly, the authors found that the susceptibility of the doped samples was governed by very large Curie tails at low temperatures, which they attributed to the dopants creating finite chains of Heisenberg spin-$\frac{1}{2}$ moments. Although no SP transition or impurity-induced order was reported for the doped samples in this study, additional measurements are needed to definitively address these questions. 

Lightly-doped Ti$_{1-x}$Sc$_{x}$OCl systems (x = 0.01 and 0.03) were studied by x-ray scattering very recently\cite{08_clancy} in an attempt to carefully address whether these systems were subject to a SP transition. These measurements confirmed that Sc-doping prevents the formation of a long-range SP state down to 7~K even at the x = 0.01 doping level and instead detect an incommensurate, short-range SP state for all temperatures below T$_{c2}$. 

The second issue of impurity-induced magnetic order can be readily addressed by the local probe technique muon spin relaxation ($\mu$SR). Due to the large gyromagnetic ratio of the muon, $\mu$SR is an extremely sensitive probe of magnetism and can readily detect internal magnetic fields as small as $\sim$~0.1~G. At TRIUMF, the muons are implanted into the sample one at a time. The muon spin precesses around the local magnetic field and then the muon decays into a positron, which is preferentially ejected along the direction of the muon spin at the time of decay (two neutrinos are also produced in the muon decay process but not detected). The $\mu$SR method is described in more detail in Ref.~\cite{99_blundell}, and previous $\mu$SR studies have confirmed the presence of antiferromagnetic order in the series of doped spin-Peierls compounds Cu$_{1-x}$Zn$_x$Ge$_{1-y}$Si$_y$O$_3$ with y as low as 0.007 and x as low as 0.021\cite{97_kojima}.

We have performed both zero-field (ZF)-$\mu$SR and longitudinal-field (LF)-$\mu$SR in this work on lightly-doped samples of Ti$_{1-x}$Sc$_x$OCl (x~$=$~0, 0.01, and 0.03). In contrast to doped CuGeO$_3$, we find no evidence for magnetic order down to 1.7 K in any of the samples investigated. Our results instead indicate that the ground state remains non-magnetic at these low doping levels.

\section{\label{sec:level2}Experimental Details}
Polycrystalline samples of TiOCl, Ti$_{0.99}$Sc$_{0.01}$OCl and Ti$_{0.97}$Sc$_{0.03}$OCl were prepared using the chemical vapor transport method and the Sc doping concentrations were inferred from susceptibility measurements as described in Ref.~\cite{03_seidel}. We performed zero field (ZF) $\mu$SR measurements on these samples to verify the existence of a non-magnetic ground state in TiOCl and to search for any evidence of magnetic ordering in the doped systems. We also performed longitudinal field (LF) $\mu$SR measurements so that the observed relaxation could be attributed to a static or dynamic mechanism. These measurements were conducted on the M20 surface muon channel at TRIUMF, using a helium gas flow cryostat in the temperature range 1.7 K $<$ T $<$ 150 K with the samples mounted in a low-background spectrometer.

\section{\label{sec:level3}Discussion and Analysis}

\begin{figure}
\begin{center}
\scalebox{0.3}{\includegraphics{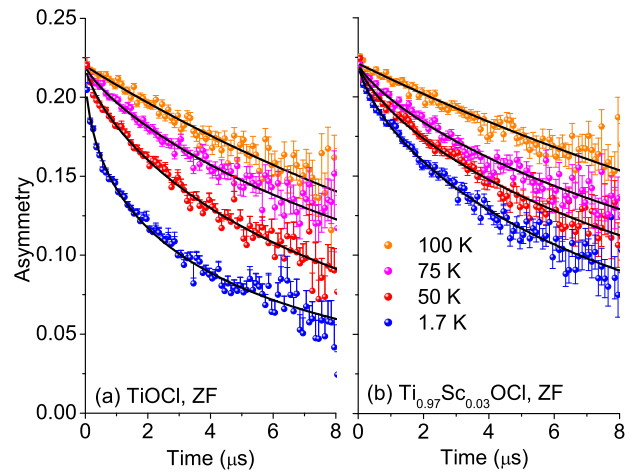}}
\caption{\label{spectra} ZF-$\mu$SR spectra of (a) TiOCl and (b) Ti$_{0.97}$Sc$_{0.03}$OCl measured at selected temperatures. The solid lines are fits to the functional form described in the text.}
\end{center}
\end{figure}

In systems with spin-singlet ground states, one expects to observe a ZF-$\mu$SR signal characteristic of a non-magnetic state: namely, the relaxation in the singlet regime should be small. This is true for the inorganic SP system CuGeO$_3$\cite{95_garcia}. In TiOCl, previous ZF-$\mu$SR measurements showed a small, gradual increase in the relaxation rate below T$_{c2}$, and then a much sharper increase in the relaxation rate below T$_{c1}$\cite{07_baker}; our results presented in Fig.~\ref{spectra}(a) are consistent with those observations. However, unlike previous work\cite{07_baker} we find no evidence that the relaxation rate saturates at low temperatures. 

One possible relaxation mechanism in TiOCl may be the slowing down of a small concentration of quasi-free spins that are created from defects/impurities. A second contribution may be the result of a muon-induced effect. If the muon site lies near a Ti-Ti dimer, this may have the effect of perturbing the local environment and creating quasi-free spins in close proximity to the muon. These quasi-free spins can then slow down and/or freeze, enhancing the relaxation rate at low temperatures. This effect has been observed in other singlet systems such as SrCu$_2$(BO$_3$)$_2$\cite{08_aczel} and KCuCl$_3$\cite{00_andreica}. To take these possible relaxation mechanisms into account in the present work, the ZF-$\mu$SR data for TiOCl was fit to the following function:
\begin{equation}
P(t)=A_0e^{-(\lambda t)^\beta}
\end{equation}
Note that the power $\beta$ was fixed to 1 above T$_{c2}$ to prevent this value from trading off with the relaxation rate as often happens when the latter value is small.

Some selected ZF-$\mu$SR spectra for Ti$_{0.97}$Sc$_{0.03}$OCl are depicted in Fig.~\ref{spectra}(b). The ZF-$\mu$SR spectra for Ti$_{0.99}$Sc$_{0.01}$OCl, while not shown explicitly, are qualitatively similar to those of the x~$=$~0.03 sample. The absence of coherent muon precession and the lack of missing asymmetry at early times indicates there is no long-range magnetic order in either of the doped materials. In light of this, the ZF-$\mu$SR data for these samples was also fit to Eq.~(1). 

\begin{figure}
\begin{center}
\scalebox{0.3}{\includegraphics{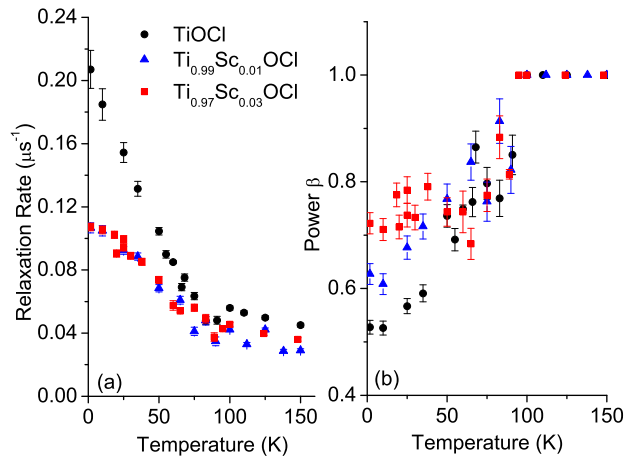}}
\caption{\label{rlx} ZF relaxation rates and $\beta$ values of Ti$_{1-x}$Sc$_{x}$OCl (x = 0, 0.01, and 0.03). Note that $\beta$ was fixed to 1 above T$_{c2}$.}
\end{center}
\end{figure}

The ZF relaxation rates and the $\beta$ values for all three systems are depicted in Fig.~\ref{rlx}. Note that TiOCl is best described by a root exponential relaxation function at the base temperature of 1.7~K, as expected for a magnetically-dilute system with rapidly fluctuating spins\cite{book_uemura}. This behaviour has also been observed in other singlet systems including Y$_2$BaNiO$_5$\cite{95_kojima} and Sr$_2$Cu$_4$O$_6$\cite{95_kojima_2}. However, the power $\beta$ increases with the doping level at the lowest temperatures investigated. One possible explanation for this behaviour is that the effective spin density is increasing in the Sc-doped cases much more than one would expect on the basis of introducing a small amount of extra impurities into the system. At these low doping levels, the deviation from root exponential behaviour should be minimal assuming the physics of the ground state hasn't changed. The large increase in $\beta$ for the doped samples may then indicate a large increase in the number of rapidly fluctuating, quasi-free spins as compared to the pure case. This result is consistent with recent x-ray work that determined the long-range, commensurate SP state is replaced by a short-range, incommensurate SP state for doping levels as low as x~$=$~0.01\cite{08_clancy}. 

\begin{figure}
\begin{center}
\scalebox{0.3}{\includegraphics{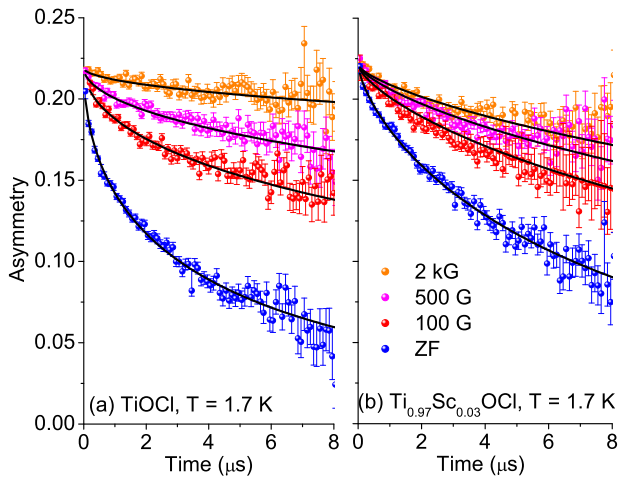}}
\caption{\label{LF_spectra} LF-$\mu$SR spectra of (a) TiOCl and (b) Ti$_{0.97}$Sc$_{0.03}$OCl for selected LF at 1.7 K. The solid lines are fits to the functional form described in the text.}
\end{center}
\end{figure}

ZF-$\mu$SR relaxation can in general be the result of static or dynamic processes.  To distinguish these two cases one needs to employ LF-$\mu$SR measurements.  If the ZF relaxation were the result of quasi-static magnetic fields, the relaxation would be decoupled in the presence of a moderate applied longitudinal magnetic field, whereas dynamic (T$_1$) relaxation would persist to much larger applied fields. Fig.~\ref{LF_spectra} shows LF-$\mu$SR data for both TiOCl and Ti$_{0.97}$Sc$_{0.03}$OCl collected at 1.7~K. Assuming a static field distribution to account for the ZF-relaxation, we obtain an estimate for the magnitude of the average internal field with the relation: B$_{loc}\sim\lambda$/$\gamma_{\mu}$ where $\gamma_{\mu}$ is the muon gyromagnetic ratio. In the cases of TiOCl and Ti$_{0.97}$Sc$_{0.03}$OCl, following this procedure leads to static field estimates of $\sim$ 2.5 and 1 G respectively. An applied LF of up to one order of magnitude greater should then be enough to completely decouple the ZF spectra. However, both TiOCl and Ti$_{0.97}$Sc$_{0.03}$OCl exhibit significant relaxation even with an applied LF of 500 G, and therefore the observed ZF relaxation must be dynamic in origin. This rules out spin freezing in these systems, especially when coupled with the lack of a characteristic peak in the ZF relaxation vs. temperature plots. 

\begin{figure}
\begin{center}
\scalebox{0.27}{\includegraphics{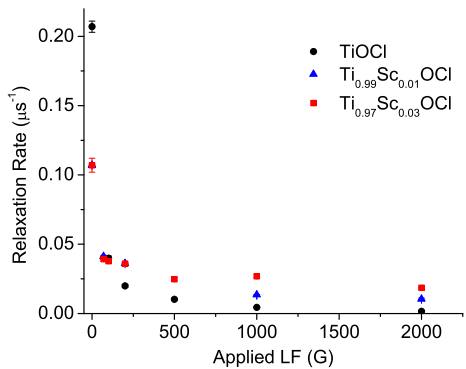}}
\caption{\label{LF_rlx} Relaxation rate as a function of applied LF for Ti$_{1-x}$Sc$_{x}$OCl (x = 0, 0.01, and 0.03) at T~$=$~1.7~K.}
\end{center}
\end{figure}

Furthermore, although the increase in $\beta$ can also be explained by the spin fluctuations of the systems slowing down with increasing x, the LF-$\mu$SR measurements rule out this possibility. The increased difficulty in completely decoupling the ZF relaxation (i.e. applying a large enough LF so the relaxation of the asymmetry effectively becomes zero) for the doped samples is quite evident. This was quantitatively characterized by fitting the LF data to Eq.~(1); $\beta$ was fixed to the ZF value for each sample. Fig.~\ref{LF_rlx} displays the resulting relaxation rates as a function of applied LF. There is some residual relaxation remaining even for the highest applied LFs in the doped samples, possibly indicating that the spin fluctuations are actually getting faster with increasing x instead and suggesting that the doped samples remain in the fast fluctuation regime. The decrease in the relaxation rate of the doped samples as compared to the pure case may then be a consequence of a motional narrowing effect.

Combined $\mu$SR, susceptibility\cite{03_seidel}, and x-ray diffraction\cite{08_clancy} results have now determined that the Ti$_{1-x}$Sc$_{x}$OCl and doped CuGeO$_3$ phase diagrams are drastically different. A commensurate, long range SP state gives way to an incommensurate, short range SP state at x~$\le$~0.01, although it is currently unknown whether there is a critical concentration for this phenomenon. This feature is accompanied by the absence of magnetic order for x~$\le$~0.03, in contrast to doped CuGeO$_3$ where impurity-induced magnetic order generally seems to persist down to very low doping concentrations. This ``magnetic order by disorder perturbation" effect has actually been proposed as a universal feature of quasi-1D spin gap systems, as impurity-induced order was also observed in the two-leg ladder system SrCu$_{2-x}$Zn$_x$O$_{3}$\cite{97_azuma}, the spin dimer system Pb$_{2-x}$(Bi, Sr)$_x$V$_3$O$_9$\cite{06_waki}, and the Haldane gap system PbNi$_{2-x}$Mg$_x$V$_2$O$_8$\cite{99_uchiyama}. Significant interchain coupling is an essential requirement for this magnetic order; the uncompensated spins resulting from doping need to be coupled in 3D space. However, the interchain interaction in TiOCl leads to frustration\cite{05_ruckamp} and it has been suggested that it is responsible for the unconventional SP behaviour found in this system. The frustrating interchain interaction may prevent the formation of antiferromagnetic long-range order upon doping TiOCl with non-magnetic Sc$^{3+}$. 

One other possible reason for the absence of magnetic order in lightly-doped Ti$_{1-x}$Sc$_{x}$OCl stems from the effects of dopant size. The ionic radius of Sc$^{3+}$ (0.745~\AA) is significantly larger than that of Ti$^{3+}$ (0.67~\AA), and this size difference may lead to local strain and lattice distortions that prevent the formation of a magnetically-ordered state. This was previously found in Cu$_{1-x}$Cd$_x$GeO$_3$\cite{98_lumsden}, where the dopant ion is also much larger than the host. For this reason, further studies on TiOCl using smaller dopant ions are necessary to help determine whether impurity-induced order is a universal feature of SP compounds. 

\section{\label{sec:level4}Conclusion}
ZF and LF-$\mu$SR measurements have verified that the ground state of TiOCl is non-magnetic, and reveal the absence of magnetic ordering and spin freezing in Ti$_{1-x}$Sc$_x$OCl (x = 0.01 and 0.03) down to 1.7 K. The latter result is in sharp contrast to the impurity-induced antiferromagnetic order observed in the other inorganic spin-Peierls system CuGeO$_3$ and many other quasi-1D spin gap systems. The difference may be due to the frustrating interchain interaction of TiOCl or the use of a dopant ion with a significantly larger ionic radius than the host. 

\begin{acknowledgments}
We acknowledge useful discussions with J.P. Clancy and B.D. Gaulin, and we appreciate the hospitality of the TRIUMF Center for Molecular and Materials Science where the $\mu$SR experiments were performed. Research at McMaster University is supported by NSERC and CIFAR. Work at Oak Ridge National Laboratory's Spallation Neutron Source was sponsored by the Scientific User Facilities Division, Office of Basic Energy Sciences, US Department of Energy. 
\end{acknowledgments}


\begin{thebibliography}{99}
\bibitem{05_lake}B.~Lake, D.A.~Tennant, C.D.~Frost, and S.E.~Nagler, Nature Materials {\bf 4}, 329 (2005).
\bibitem{99_kageyama}H.~Kageyama, K.~Yoshimura, R.~Stern, N.V.~Mushnikov, K.~Onizuka, M.~Kato, K.~Kosuge, C.P.~Slichter, T.~Goto, and Y.~Ueda, Phys. Rev. Lett. {\bf 82}, 3168 (1999). 
\bibitem{96_dagotto}E.~Dagotto and T.M.~Rice, Science {\bf 271}, 618 (1996).
\bibitem{10_hsu}C.H.~Hsu, J.-Y.~Lin, W.L.~Lee, M.-W.~Chu, T.~Imai, Y.J.~Kao, C.D.~Hu, H.L.~Liu, and F.C.~Chou, Phys. Rev. B {\bf 82}, 094450 (2010).
\bibitem{93_hase}M. Hase, I. Terasaki, and K. Uchinokura, Phys. Rev. Lett. {\bf 70}, 3651 (1993).
\bibitem{03_seidel}A. Seidel, C.A. Marianetti, F.C. Chou, G. Ceder, and P.A. Lee, Phys. Rev. B {\bf 67}, 020405(R) (2003).
\bibitem{05_sasaki}T. Sasaki, M. Mizumaki, K. Kato, Y. Watabe, Y. Nishihata, M. Takata, and J. Akimitsu, J. Phys. Soc. Japan {\bf 74}, 2185 (2005).
\bibitem{75_bray}J.W. Bray, H.R. Hart Jr., L.V. Interrante, I.S. Jacobs, J.S. Kasper, G.D. Watkins, S.H. Wee, and J.C. Bonner, Phys. Rev. Lett. {\bf 35}, 744 (1975).
\bibitem{79_huizinga}S. Huizinga, J. Kommandeur, G.A. Sawatzky, B.T. Thole, K. Kopinga, W.J.M. de Jonge, and J. Roos, Phys. Rev. B {\bf 19}, 4723 (1979).
\bibitem{03_imai}T. Imai and F.C. Chou, cond-mat/0301425 (unpublished).
\bibitem{05_shaz}M. Shaz, S. van Smaalen, L. Palatinus, M. Hoinkis, M. Klemm, S. Horn, and R. Claessen, Phys. Rev. B {\bf 71}, 100405(R) (2005).
\bibitem{07_clancy}J.P. Clancy, B.D. Gaulin, K.C. Rule, J.P. Castellan, and F.C. Chou, Phys. Rev. B {\bf 75}, 100401(R) (2007).
\bibitem{07_abel}E.T. Abel, K. Matan, F.C. Chou, E.D. Isaacs, D.E. Moncton, H. Sinn, A. Alatas, and Y.S. Lee, Phys. Rev. B {\bf 76}, 214304 (2007).
\bibitem{07_saha}S.R. Saha, S. Golin, T. Imai, and F.C. Chou, Journal of Physics and Chemistry of Solids {\bf 68}, 2044 (2007).
\bibitem{95_hase}M.~Hase, N.~Koide, K.~Manabe, Y.~Sasago, K.~Uchinokura, and A.~Sawa, Physica B {\bf 215}, 164 (1995).
\bibitem{95_oseroff}S.B. Oseroff, S-W. Cheong, B. Aktas, M.F. Hundley, Z. Fisk, and L.W. Rupp Jr., Phys. Rev. Lett. {\bf 74}, 1450 (1995).
\bibitem{99_blundell}S.J.~Blundell, Contemporary Physics {\bf 40}, 175 (1999).
\bibitem{97_kojima}K.M.~Kojima, Y.~Fudamoto, M.~Larkin, G.M.~Luke, J.~Merrin, B.~Nachumi, Y.J.~Uemura, M.~Hase, Y.~Sasago, K.~Uchinokura, Y.~Ajiro, A.~Revcolevschi, and J.-P.~Renard, Phys. Rev. Lett. {\bf 79}, 503 (1997).
\bibitem{97_martin}M.C. Martin, M. Hase, K. Hirota, G. Shirane, Y. Sasago, N. Koide, and K. Uchinokura, Phys. Rev. B {\bf 56}, 3173 (1997).
\bibitem{98_grenier}B. Grenier, J.-P. Renard, P. Veillet, C. Paulsen, G. Dhalenne, and A. Revcolevschi, Physica B {\bf 259-261}, 954 (1999).
\bibitem{98_manabe}K. Manabe, H. Ishimoto, N. Koide, Y. Sasago, and K. Uchinokura, Phys. Rev. B {\bf 58}, R575 (1998).
\bibitem{07_haravifard}S. Haravifard, K.C. Rule, H.A. Dabkowska, B.D. Gaulin, Z. Yamani, and W.J.L. Buyers, Journal of Physics: Condensed Matter {\bf 19}, 436222 (2007).
\bibitem{98_lumsden} M.D.~Lumsden, B.D.~Gaulin, and H.~Dabkowska, Phys. Rev. B {\bf 58}, 12252 (1998).
\bibitem{97_anderson}P.E. Anderson, J.Z. Liu, and R.N. Shelton, Phys. Rev. B {\bf 56}, 11014 (1997).
\bibitem{02_uchinokura}K.~Uchinokura, J. Phys. Cond. Matt. {\bf 14}, R195 (2002).
\bibitem{96_fukuyama}H.~Fukuyama, T.~Tanimoto, and M.~Saito, J. Phys. Soc. Japan {\bf 65}, 1182 (1996). 
\bibitem{08_clancy}J.P. Clancy, B.D. Gaulin, J.P. Castellan, K.C. Rule, and F.C. Chou, Phys. Rev. B {\bf 78}, 014433 (2008).
\bibitem{95_garcia}J.L.~Garcia-Munoz, M.~Suaaidi, and B.~Martinez, Phys. Rev. B {\bf 52}, 4288 (1995).
\bibitem{07_baker}P.J.~Baker, S.J.~Blundell, F.L.~Pratt, T.~Lancaster, M.L.~Brooks, W.~Hayes, M.~Isobe, Y.~Ueda, M.~Hoinkis, M.~Sing, M.~Klemm, S.~Horn, and R.~Klaasen, Phys. Rev. B. {\bf 75}, 094404 (2007).
\bibitem{08_aczel}A.A. Aczel, G.J. MacDougall, J.A. Rodriguez, G.M. Luke, P.L. Russo, A.T. Savici, Y.J. Uemura, H.A. Dabkowska, C.R. Wiebe, J.A. Janik, and H. Kageyama, Phys. Rev. B {\bf 76}, 214427 (2007).
\bibitem{00_andreica}D.~Andreica, N.~Cavadini, H.U.~Gudel, F.N.~Gygax, K.~Kramer,  M.~Pinkpank and A.~Schenck, Physica B {\bf 289}, 176 (2000).
\bibitem{book_uemura}Y.J. Uemura in $\mu$SR Relaxation Functions in Muon Science, Edited by S.L. Lee, S.H. Kilcoyne, and R. Cywinski, IOP Publishing, Edinburgh and Briston UK (1999).
\bibitem{95_kojima}K.M.~Kojima, A. Keren, G.M. Luke, B. Nachumi, W.D. Wu, Y.J. Uemura, M. Azuma, and M. Takano, Phys. Rev. Lett. {\bf 74}, 2812 (1995).
\bibitem{95_kojima_2}K.M.~Kojima, A. Keren, L.P. Le, G.M. Luke, B. Nachumi, W.D. Wu, Y.J. Uemura, K. Kiyono, S. Miyasaka, H. Takagi, and S. Uchida, Phys. Rev. Lett. {\bf 74}, 3471 (1995).
\bibitem{97_azuma}M. Azuma, Y. Fujishiro, M. Takano, M. Nohara, and H. Takagi, Phys. Rev. B {\bf 55}, R8658 (1997).
\bibitem{06_waki}T.~Waki, Y.~Itoh, C.~Michioka, K.~Yoshimura, and M.~Kato, Phys. Rev. B {\bf 73}, 064419 (2006).
\bibitem{99_uchiyama}Y.~Uchiyama, Y.~Sasago, I.~Tsukada, K.~Uchinokura, A.~Zheludev, T.~Hayashi, N.~Miura, and P.~Boni, Phys. Rev. Lett. {\bf 83}, 632 (1999).
\bibitem{05_ruckamp}R. Ruckamp, J. Baier, M. Kriener, M.W. Haverkort, T. Lorenz, G.S. Uhrig, L. Jongen, A. Moller, G. Meyer, and M. Gruninger, Phys. Rev. Lett. {\bf 95}, 097203 (2005).
\end{thebibliography}
\end{document}